\pdfoutput=1

\documentclass[journal]{IEEEtran}
\ifCLASSINFOpdf
  \usepackage[pdftex]{graphicx}
  \graphicspath{{../pdf/}{../jpeg/}}
  \DeclareGraphicsExtensions{.pdf,.jpeg,.png}
\else
  \usepackage[dvips]{graphicx}
  \graphicspath{{../eps/}}
  \DeclareGraphicsExtensions{.eps}
\fi

\usepackage{amssymb}
\usepackage{graphicx}
\usepackage{subfigure}
\usepackage{booktabs}
\usepackage{threeparttable}
\usepackage{amsmath}
\usepackage{color}

\usepackage{subfigure}
\usepackage[linesnumbered, boxed, ruled, commentsnumbered]{algorithm2e}
\usepackage{multirow}
\begin{document}
%
\title{MLFcGAN: Multi-level Feature Fusion based Conditional GAN for Underwater Image Color Correction}

\author{Xiaodong Liu , Zhi Gao and Ben M. Chen
\thanks{ Corresponding authors: Xiaodong Liu ; Zhi Gao.
}
\thanks{
X. Liu is with the  Department of Electrical and Computer Engineering, National University of Singapore , 117583 Singapore. (e-mail: xiaodongliu@u.nus.edu).}
\thanks{
	Z. Gao is with the the School of Remote Sensing and Information
	Engineering, Wuhan University, Wuhan 430079, China. (e-mail: gaozhinus@gmail.com).}
\thanks{B. M. Chen  is with the Department of Mechanical and Automation Engineering, Chinese University of Hong Kong, Shatin, N.T., Hong Kong; and the Department of Electrical and Computer Engineering, National University of Singapore, Singapore 117583.(e-mail: bmchen@cuhk.edu.hk; bmchen@nus.edu.sg ). }

}

%
%

\maketitle

\begin{abstract}
Color correction  for underwater images has received increasing interests, due to its critical role in facilitating available mature vision algorithms for underwater scenarios. Inspired by the  stunning success of deep convolutional neural networks (DCNNs) techniques in many vision tasks, especially the strength in extracting features in multiple scales, we propose a deep multi-scale feature fusion net based on the conditional generative adversarial network (GAN) for underwater image color correction. In our network, multi-scale features are extracted first, followed by augmenting local features in each scale with global features. This design was verified to facilitate more effective and faster network learning, resulting in better performance in both color correction and detail preservation. We conducted extensive experiments and compared with the state-of-the-arts approaches quantitatively and qualitatively, showing that our method achieves significant improvements. 
\end{abstract}


%
\IEEEpeerreviewmaketitle

\section{Introduction}
\label{intro}

Underwater imaging has been proven valuable in numerous remote sensing applications \cite{matteoli2014automated}, \cite{fandos2013unified}, where remote sensors such as sonar, LIDAR are deployed conventionally. Due to the recent advances in both hardware and algorithmic approaches, affordable and compact off-the-shelf underwater cameras are becoming popular, allowing people easily to collect images from a wide range of undersea world by either divers or remotely operated submersibles. These captured underwater images and videos  with color  information are valuable resources for many underwater scientific remote sensing missions, such as the marine biology \cite{ludvigsen2007applications} and ecological  research \cite{strachan1993recognition}.

Compared with everyday images captured in air, underwater images typically suffer from color shift and relatively low quality due to the light absorption and the light scattering, posing significant challenges to available mature vision algorithms in achieving expected performance. 
For underwater scenes, light absorption is wavelength dependent, the longer the wavelength, the higher the absorption rate is. Thus, the red component of light is absorbed first, and underwater images often appear bluish or greenish. Severe underwater light scattering is due to the relatively larger particles than those in air, resulting in decreased visibility. Moreover, such absorption and scattering effects are  hard to be explicitly modeled, as they are relevant to many factors, such as water temperature, salinity, and types of particles, etc. This complex degradation makes it challenging to restore the visibility and color of underwater images \cite{lu2017underwater}. Restoring underwater images with natural colors and fine details still remains an open problem \cite{shortis2016review}. 
Several pioneering works tackle this issue by inferencing the non-degraded images based on the  image degradation model.  As this inverse problem is ill-posed,  prior knowledge or assumptions are introduced to obtain an solution. These include the approaches based on dark channel prior \cite{he2011single} and its variants \cite{chiang2012underwater}, methods based on haze-lines prior \cite{berman2018underwater} etc. However, the prior knowledge may fail for some underwater scenes and  all these model-based methods reported less competitive image color correction results \cite{wang2017deep}. \\ 
Like many other computer vision tasks, underwater image color correction has been benefiting from the deep convolutional neural networks (DCNNs). In \cite{wang2017deep}, the  convolutional neural network (CNN) was trained to approximate the underwater image restoration function given the synthesized paired underwater images. Li et al.  in   \cite{li2018watergan} proposed a two-stage CNN for depth estimation and color restoration. 
Recently, the generative adversarial networks have achieved huge success on many tasks such as super-resolution \cite{jiang2019edge} and image synthesis and translation \cite{isola2017image}. Inspired by this,
in  \cite{fabbri2018enhancing}, the conditional generative adversarial network (cGAN) was exploited to address the  underwater image enhancement as the image-to-image translation problem.  Based on \cite{fabbri2018enhancing},  the authors in \cite{yu2018underwater} introduced perceptual loss into cGAN framework for underwater image color correction. Later, CycleGAN was introduced for color correction in \cite{li2018emerging} and \cite{lu2019multi}.
Generally speaking, such CNN-based methods outperform aforementioned model-based methods. However, methods in \cite{wang2017deep}, \cite{li2018watergan} and \cite{fabbri2018enhancing} only leverage low-level local features with relatively shallow networks and such low-level features captured with limited receptive field can hardly encode the high-level semantic knowledge, resulting in noisy and imperfect color restoration results.
To overcome the limitations of available CNN-based methods, we propose to exploit high-level features in the cGAN framework for underwater image color correction. Leveraging on the high-level features, impressive improvements have been made for detection  \cite{dai2016instance}, segmentation \cite{long2015fully}, and pose estimation \cite{yang2016end} etc.
Different to those methods, we augment local features of each level with global features that capture the semantic information, such as the overall lighting condition, and scene layout of the whole image,  for underwater image color correction. \\ 
In this paper,  we propose a generic multi-level features fusion framework based on conditional GANs (MLFcGAN) for underwater image color correction. Compared to the existing network structures, MLFcGAN extracts more scales of features and the global features are fused with low-level features at each scale. Extensive experiments are conducted and comparisons with the state-of-the-arts approaches are made quantitatively and qualitatively, showing that our design achieves significant improvements.

\begin{figure}[!t]
	\centering
	\includegraphics[width=\linewidth]{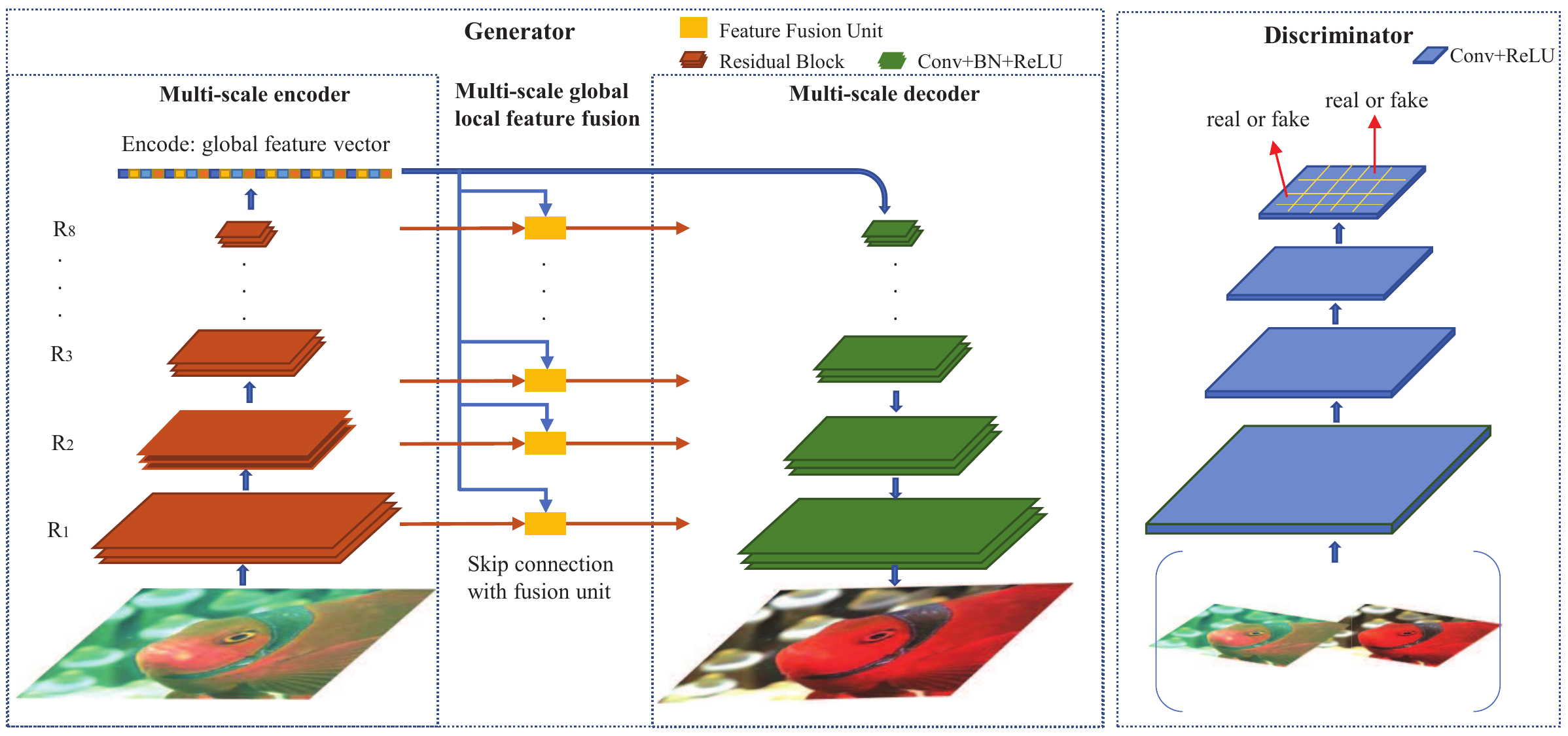}
	\caption{Overview of the network structure. On the left is the generator which consists of the encoder, decoder and multi-scale local and global features fusion with skip connection. On the right is the PatchGAN based discriminator.}
	\label{fig:network_structure. }
\end{figure}

\section{Proposed Method}
\label{section:ournetwork} 
Our design is based on the framework of condition GAN where the adversarial loss is beneficial to image generation compared to DCNNs with Euclidean loss \cite{isola2017image}. Consisting of the one generator $ G $ and one discriminator $ D $, GAN is initially deployed to produce vivid images given noise input $ z $.   $ G $ aims to produce images to fool $ D $, and $ D $ is trained to distinguish samples from real images, where $ G $ and $ D $ are updated in an adversarial fashion.  Slightly different from the basic GAN, conditional GAN takes conditional variables as inputs.

\subsection{The Generator}
The generator is based on the encode-decoder structure and multi-scale feature extraction and feature fusion unit are designed and novelly integrated, whose effectiveness is demonstrated in section \ref{sec:ablation_study}.
\subsubsection{ Global features and multi-level local features extraction}
High-level information extracted from the image with a receptive field of the entire image is termed as the global features. The extraction of  this information is prevalent in feature engineering, and can be achieved by the average pooling along the spacial  dimension as described in \cite{hu2018squeeze}.  Unlike the average pooling, we extract the global features by gradually down-sampling the input images with convolution layers until the output has a dimension $ 1\times1 \times c_g  $, where  $ c_g $ represents the number of channels of global features.  The benefits of gradually down-sampling compared to simple average pooling are two folds: First, the number of feature maps at each resolution are free to be chosen. Second, in this way,  more  scales of local features can be extracted simultaneously. The multi-scale local features offer abstraction of the image at different resolution, and are beneficial for generating images with fine details. (This is validated in section \ref{sec:ablation_study}.)  As the encoder part of the generator functions as feature extraction, and to make the network deeper and easier to optimize, the  residual building blocks are employed as proposed in \cite{He2016DeepRL}.
\subsubsection{Fusion of the global and local features unit}\label{sec:fustion}
Here, we propose the feature fusion unit to dynamically fuse the global features with local features. Suppose the local feature map $ f_l $ at scale $ i $ with dimension $ h_i \times w_i \times c_i $ and the global features $ f_{g} $ with dimension $ 1 \times 1 \times c_g $. The global features  are first adjusted through a $ 1\times1 $ convolution layer with learnable weights for channel matching. Normally, $ c_g $ is larger than $ c_i $, therefore, this step is designed to adaptively extract the most useful information from global features for local features at scale $ i $.
The parameter settings for this convolution layer are: kernel size $ 1\times 1 $, stride $ 1 $, number of input channels $ c_g $ and number of output channels $ c_i $.

Denote $ F_{conv} $, $ F_{copy} $, $ F_{reshape} $, $ F_{concat} $  as the convolution, copy, reshape and concatenate operation separately. The feature fusion unit follows the following operations as described in equations (\ref{eq:con}), (\ref{eq:copy}), (\ref{eq:reshape}), (\ref{eq:concat}).
The output  after the convolution can be denoted as:
\begin{equation}\label{eq:con}
f_{g1} = F_{conv}(f_g, W)
\end{equation}
where $ W $ denotes the learnable weights.
Then $ f_{g1} $ is copied in total $ h_i \times w_i $ times:
\begin{equation}\label{eq:copy}
f_{g2} = F_{copy}(f_{g1}, num=h_i \times w_i  )
\end{equation}
then $ f_{g2} $ are reshaped into  $ (h_i , w_i, c_i) $: 
\begin{equation}\label{eq:reshape}
f_{g3} = F_{reshape}(f_{g2},size=(h_i, w_i, c_i))
\end{equation}
Lastly the feature $ f_l $ and $ f_{g3} $ with the same dimension are concatenated along the channel dimension:
\begin{equation}\label{eq:concat}
f_{out} = F_{concat}(f_{l},f_{g3})
\end{equation}
The fused feature $ f_{out} $ are fed into the corresponding layer in the decoder with skip connection. 

\textit{Remarks:} Local features and global features cover different scales of the image and convey variant knowledge of the image. Normally, the local image features represent low-level features such as edges. The global features encode the  high-level information such as the overall light condition, the layout or the type of the scene etc. Fusing with low and high information at different scales is beneficial to generate images with plausible natural color and better details. In addition, since the global features are a  higher abstraction of local features, they could act as the regularizer to penalize the artifacts generated in the enhanced images due to the mishandling in the low resolution images. Hence, here, we fuse the global features with low-level features at each resolution as shown in figure \ref{fig:network_structure. }.
\subsection{The Discriminator}  
 In this paper, PatchGAN is adopted as the discriminator  \cite{isola2017image}.
PatchGAN is designed to identify if each $ N\times N $ patch in the image is real or generated by $ G $, and the overall decision is achieved by averaging the authenticity of all patches. In this case, the generated image will only be considered as real, when nearly all image patches are generated with good and less blurring details to be considered with high probability to be real. Considering the network is trained in an adversarial way, the generator is pushed to focus more on high frequency (details in the image) and generate better image details so as to fool  the  discriminator. On the other hand, compared to the whole-image discriminator, less convolutional layers are needed for PatchGAN. For more details, please see \cite{isola2017image}.


\subsection{ Objective Function}
 Considering the problem of exploding or vanish gradient during training the original GAN \cite{goodfellow2014generative}\cite{yu2018underwater},  many improvements and modifications have been made to stabilize the training such as the LSGAN \cite{mao2017least}, Wasserstein GAN \cite{arjovsky2017wasserstein}, Wasserstein GAN with gradient penalty (WGAN-GP), and the WGAN-GP shows the best performance for image generation \cite{gulrajani2017improved}. In this paper, the WGAN-GP \cite{gulrajani2017improved} loss is adopted and modified into conditional setting as the adversarial loss:
\begin{equation}\label{eq:wgan loss}
\begin{array}{l}
{{\cal L}_{cWGAN-GP}} = {E_{x,y}}[D(x,y)] - {E_x}[D(x,G(x))]\\
\begin{array}{*{20}{c}}
{}&{}&{}&{}
\end{array} + \lambda {E_{\hat x}}[{(||{_{\hat x}}D(\hat x)|{|_2} - 1)^2}]
\end{array}
\end{equation} 
where $ x $, $ y $ are the original raw image and ground-truth underwater image (with good color balance and details), $ \hat{x} $ are the samples along the lines between the generated images $ G(x) $ and $ y $, and $ \lambda $ stands for the weight factor.

The adversarial loss measures the Wasserstein distance from the distribution perspective between the distribution of the generated images and that of the groud-truth images. The traditional loss such as the $ L_{2} $ or $ L_{1} $ loss measure the distance from the pixel  perspective, and it has been demonstrated  helpful to combine the adversarial loss with  traditional distance loss  for image-to-image translation tasks \cite{pathak2016context}.
As reported in \cite{zhao2015loss}, \cite{isola2017image}, $ L_{1} $ loss is likely to give less blurring results than $ L_{2} $ loss, therefore,  the $ L_1 $  loss is introduced:
\begin{equation}\label{l1loss}
\mathcal{L}_{L_1}(G)=E_{x,y}[||y-G(x)||_1]
\end{equation}  
The overall objective function $ \mathcal{L^{*}} $  is ( $ \lambda_1 $ is the weight factor ):
\begin{equation}\label{totla loss}
\mathcal{L^{*}} = \mathop {\min }\limits_G \mathop {\max }\limits_D \mathcal{L}_{cWGAN-GP}(G,D) + \lambda_1 \mathcal{L}_{L_1}(G)
\end{equation}

\section{Experiments and Analysis}

\subsection{Preparation} 
\label{subsection:prepare}
\subsubsection{Data} 
Different from  other computer vision tasks, there is limited publicly available dataset for underwater images as the challenge to acquire the ground-truth. The dataset used for training was recently proposed in  \cite{fabbri2018enhancing}. 
It contains 6,128 image pairs with ground truth (non-distorted underwater images) and distorted underwater images. We randomly select 6,000 image pairs as the training set and the remaining images are used for validation.
In addition, to further evaluate the generalization ability of our method, 86 real world underwater images are collected from the Internet with various scenes. All the images are resized to $ 256 \times 256 $.
\label{subsubsection:data}

\subsubsection{Training settings}
In our experiment, we set $ \lambda =10$  and $ \lambda_1 =10$. The values are selected based on the basic hyper-parameter tuning. We apply the Adam solver, with the learning rate = 0.0002, $ \beta_{1}=0.5 $, $ \beta_{2}=0.999 $. Batch size is 1 and the network is trained for 50 epochs.
\subsubsection{Methods for comparison} 
Comparisons are made with the following state-of-the-arts methods  which are published in top conferences or journals recently. a) \textbf{Model-based methods}: EUF \cite{ancuti2012enhancing},
MBIE \cite{cho2018model}, HLC \cite{berman2018underwater},
UVE \cite{cho2017visibility}, and they are tested with the code provided by their authors.  b) \textbf{Learning based methods}: 
UGAN \cite{fabbri2018enhancing} and DUIENet \cite{li2019underwater}, UWCNN \cite{li2020underwater} and CycleGAN \cite{zhu2017unpaired}. 
UGAN \cite{fabbri2018enhancing} and CycleGAN \cite{zhu2017unpaired} are re-trained on the dataset from  scratch with the recommended parameter settings in their papers to achieve the best enhancement results. DUIENet \cite{li2019underwater} and UWCNN \cite{li2020underwater} are tested with the pretrained model provided by the authors.


 
\subsubsection{Evaluation metrics} 
The evaluation is based on both the validation images and real world underwater images. For the validation images set where the ground truth are available, the peak signal to noise ratio (PSNR) and structural similarity index (SSIM) are adopted for quantitatively comparisons. 


\subsection{Experimental results} 
\label{subsection:results}
\begin{figure}[!h]
	\centering
	\includegraphics[width=\linewidth]{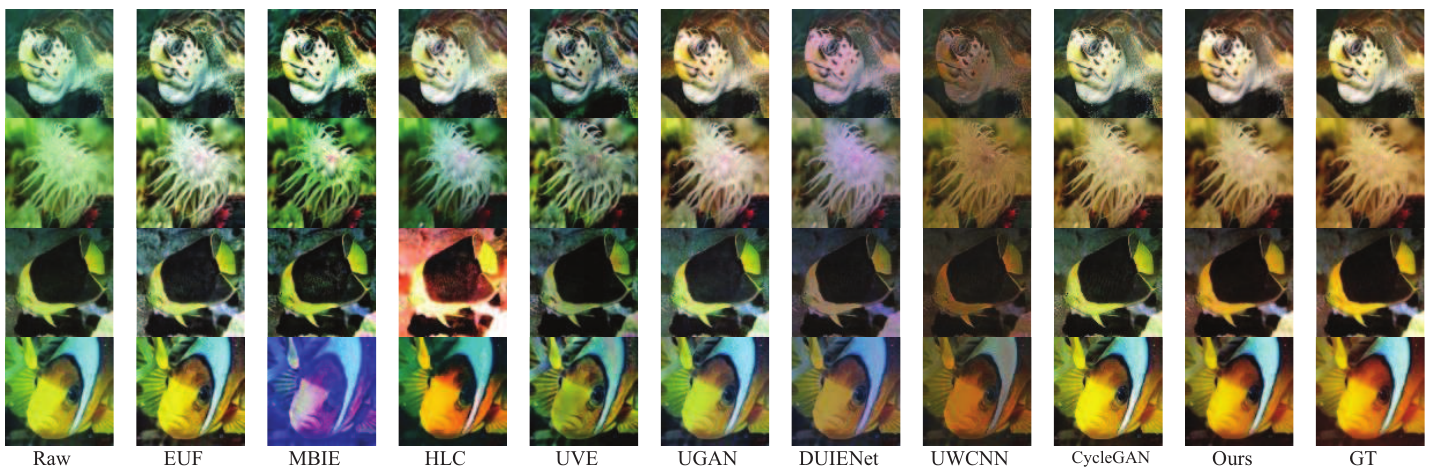}
	\caption{Color correction comparisons on \textit{val} images with EUF \cite{ancuti2012enhancing}, MBIE \cite{cho2018model}, HLC  \cite{berman2018underwater}, UVE  \cite{cho2017visibility}, UGAN  \cite{fabbri2018enhancing} and DUIENet \cite{li2019underwater}, UWCNN \cite{li2020underwater}, CycleGAN \cite{zhu2017unpaired}. Better zoom in 4x for view.}
	\label{fig:result_val_short}
\end{figure}

\textbf{Evaluation on the validation image set.}
The average PSNR and SSIM were calculated and shown in  table \ref{tab:psnr_ssim_val_images}. 
 In figure \ref{fig:result_val_short}, it demonstrates that our method can achieve the best image restoration performance which is rather close to the ground truth.
 Quantitatively, as can be seen from table \ref{tab:psnr_ssim_val_images}, our method achieves the best PSNR and SSIM and outperforms other methods with a large margin, demonstrating the superior learning ability of our structure.

\begin{table*}[h]
	\scriptsize
	\centering
	\caption{PSNR and SSIM evaluation on validation images.}
	\label{tab:psnr_ssim_val_images}
	\begin{tabular}{c|ccccccccc}\hline
		&  EUF & MBIE & HLC  & UVE    & UGAN  &  DUIENet &  UWCNN &  CycleGAN &Ours \\ \hline \hline
		PSNR           & 16.36  & 14.67  & 15.72  & 15.58  & 18.58  & 19.6279& 15.2351& 23.0224& \textbf{23.42}  \\
		SSIM           & 0.5527 & 0.4244 & 0.5249 & 0.4716 & 0.5851 & 0.6431& 0.6133& 0.7893 &\textbf{0.8158}\\ \hline  
	\end{tabular}
\end{table*}

\begin{figure}[!h]
 	\centering
	\includegraphics[width=\linewidth]{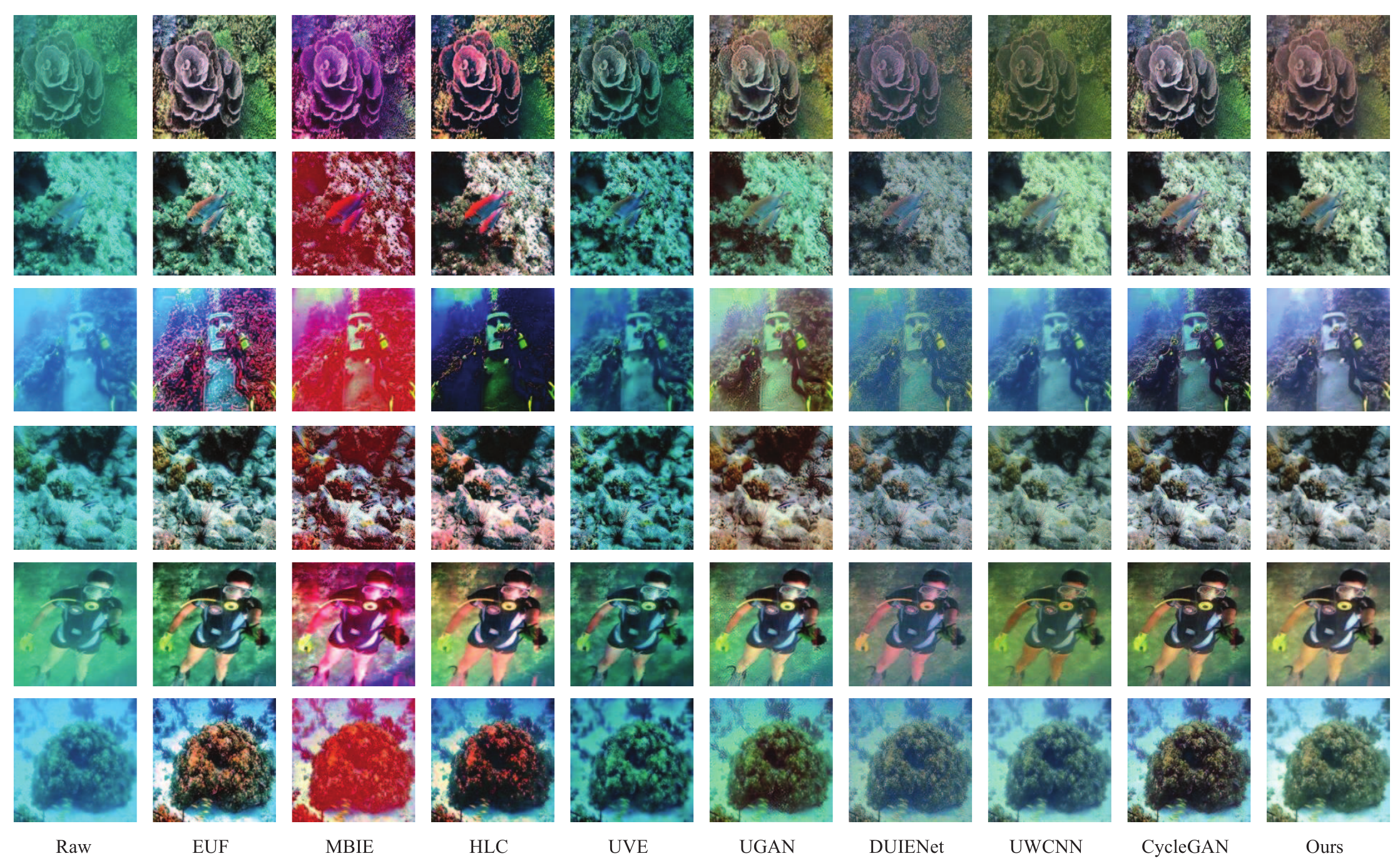}
 	\caption{Color correction results comparisons on \textit{real} underwater images with EUF \cite{ancuti2012enhancing}, MBIE \cite{cho2018model}, HLC  \cite{berman2018underwater}, UVE  \cite{cho2017visibility}, UGAN  \cite{fabbri2018enhancing} and DUIENet \cite{li2019underwater}, UWCNN \cite{li2020underwater} and  CycleGAN \cite{zhu2017unpaired}. Better zoom in for view. }
 	\label{fig:result_test_short}
 \end{figure}

\begin{figure}[!h]
 	\centering
 	\includegraphics[width=\linewidth]{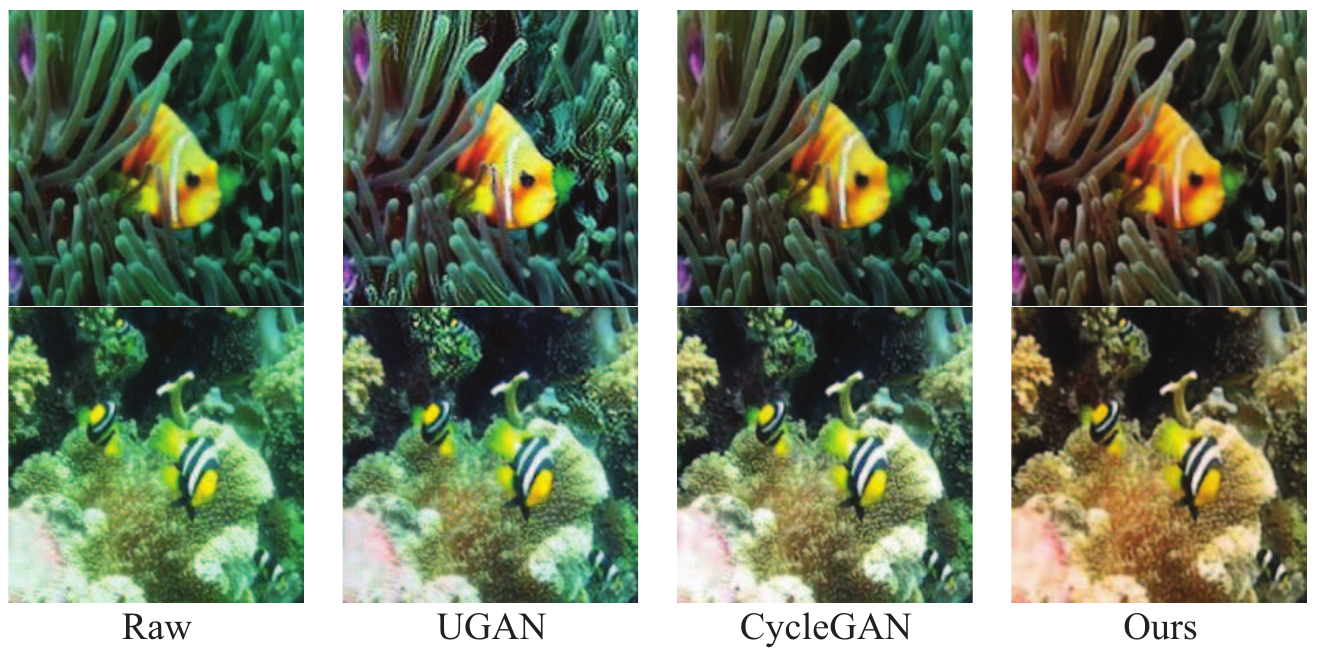}
 	\caption{Further comparisons with  UGAN  \cite{fabbri2018enhancing}, CycleGAN \cite{zhu2017unpaired}.  UGAN and CycleGAN may fail for deep  green hue scenarios and generate more blurry results than ours.}
 	\label{fig:UGAN_CYCLEGAN}
 \end{figure}

\begin{figure}[!h]
	\centering
	\includegraphics[width=\linewidth]{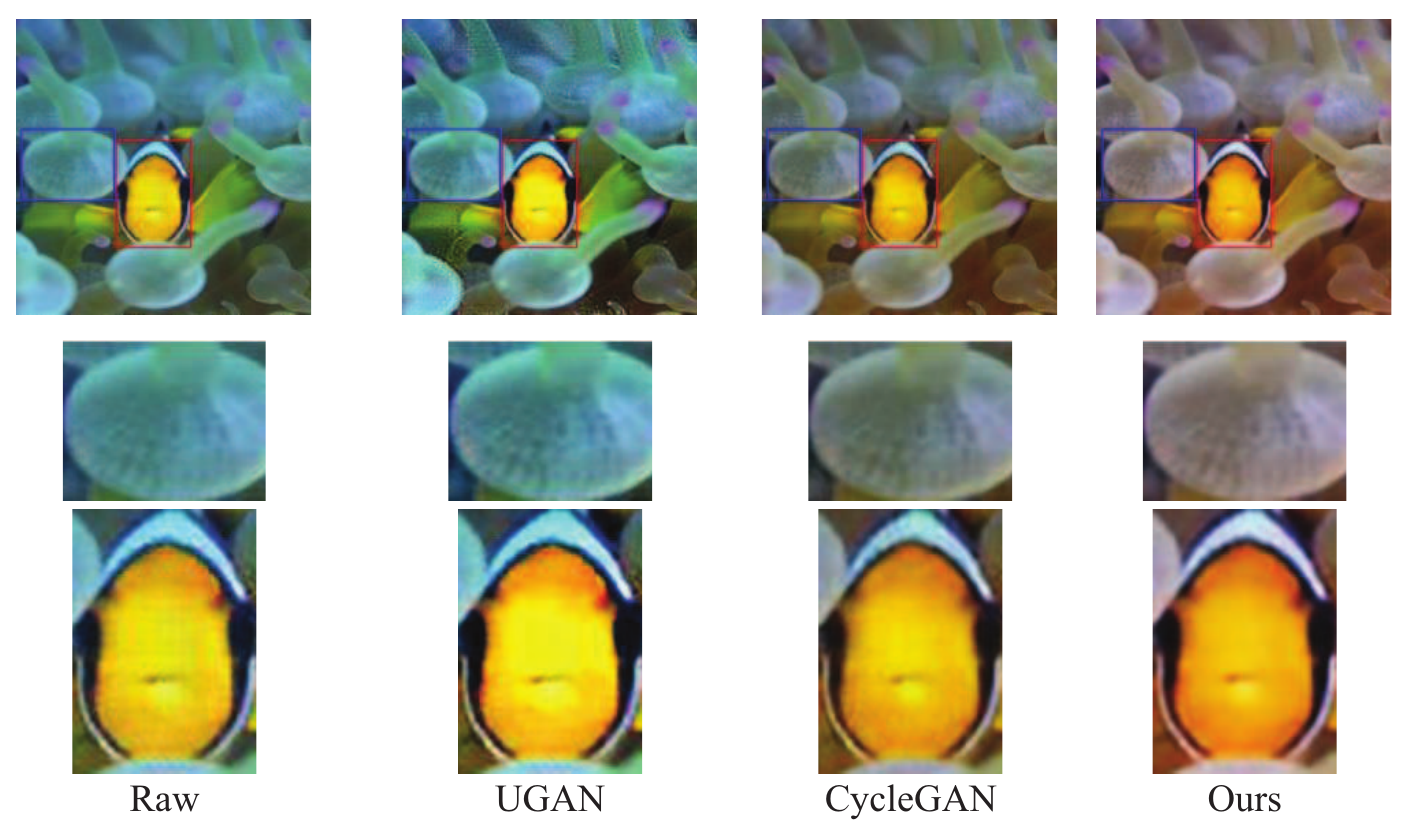}
	\caption{Zoom in comparisons on the image patches against UGAN \cite{fabbri2018enhancing} and CycleGAN \cite{zhu2017unpaired}, where they may generate more blurry results, and  our method  could restore natural color and produce smooth results without loss of details.}
	\label{fig:zoom_in_comparisons}
\end{figure}

\textbf{Evaluation on the real world underwater images.}  Visual comparisons are presented in figure \ref{fig:result_test_short}.
 UVE \cite{cho2017visibility}, HLC \cite{berman2018underwater} perform poorly to remove the color cast, and images processed by these methods remain bluish or greenish.
 MBIE \cite{cho2018model}, on the other hand, can generate images with good color saturation , but it introduces  unwanted superfluous red hue.  Similarily, EUF \cite{ancuti2012enhancing} introduces unwanted  artifacts and noise. The poor performance of the model-based methods  may  be because that the prior knowledge or the assumed parameters may not hold for some underwater scenes.  
For DUIENet \cite{li2019underwater},  it is likely to  introduces unexpected gray hue ( such as the images in the $ 2^{nd} $ and $ 4^{th} $ row) or red hue (such as the images in the $ 1^{st} $ and $ 5^{th} $ row) with lower brightness as shown in figure \ref{fig:result_test_short}. 
  For UWCNN \cite{li2020underwater}, similarly,  the enhanced results appear dim with low intensity as show in figure \ref{fig:result_val_short} and figure \ref{fig:result_test_short}.
  As UGAN and CycleGAN are re-trained on the unified dataset, we made further comparisons in figure \ref{fig:UGAN_CYCLEGAN} and figure \ref{fig:zoom_in_comparisons}.
  As for UGAN \cite{fabbri2018enhancing}, the color correction performance is generally acceptable, but it is likely to generate noisy patches in the texture-less areas and along the boundary of the images. 
   For CycleGAN \cite{zhu2017unpaired}, it may fail for some cases (such as the deep green scenarios) and retain the green hue with blurry details, as shown in figures \ref{fig:UGAN_CYCLEGAN}  and \ref{fig:zoom_in_comparisons}. On the other hand, our method can  achieve good color correction results with smooth details.
 As discussed earlier, only based on the limited scales of local features, it is likely to produce noisy patches. On the other hand, the proposed multi-scale features extraction and  fusion strategy can help  generate images with better details preservation.
 
 
\section{Generator Structure Comparisons}
\label{sec:ablation_study}
\begin{figure}[!t]
	\centering     
	\subfigure[PSNR]{\label{fig:generator_test_psnr}\includegraphics[width=0.49\linewidth]{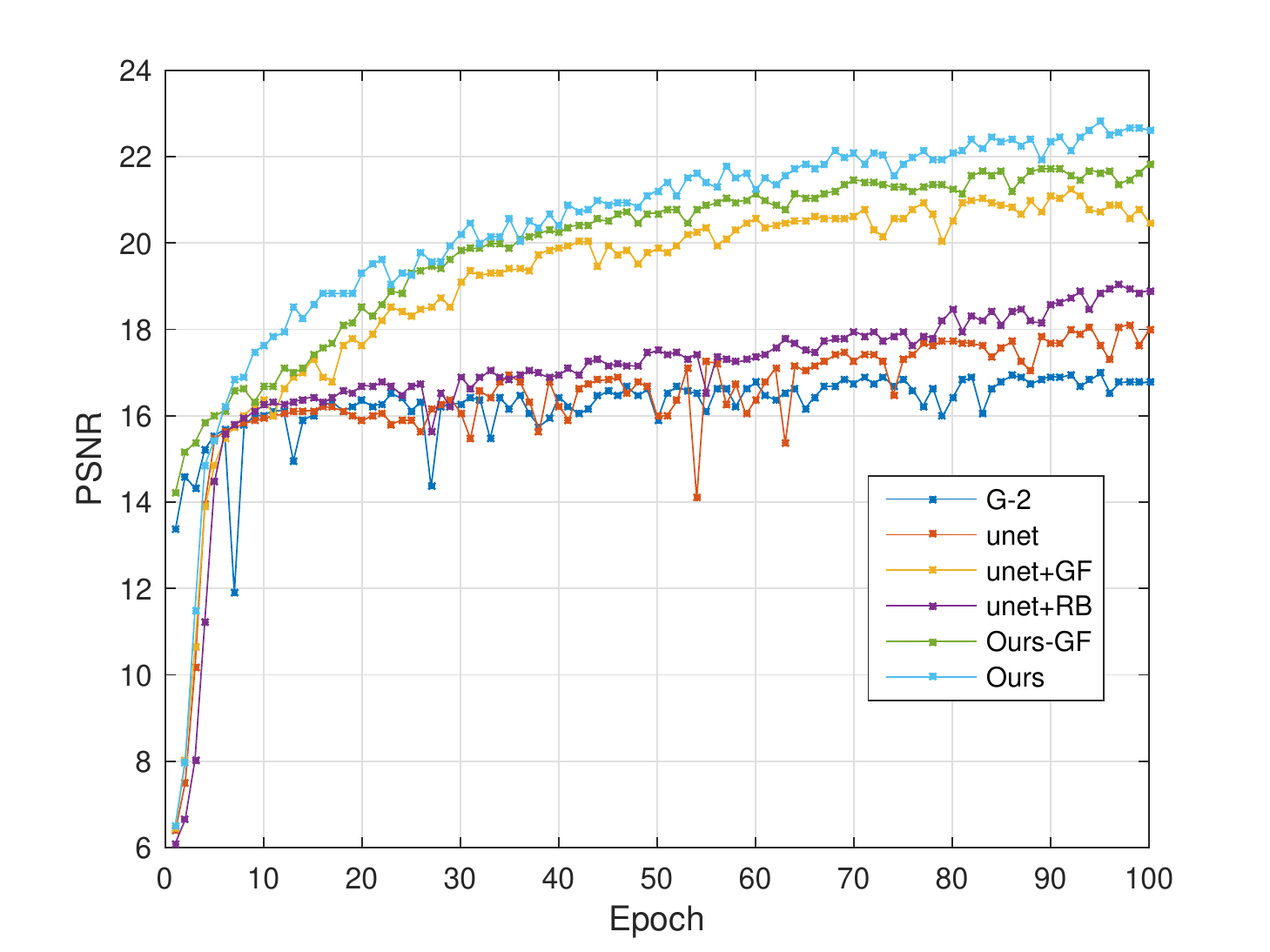}}
	\subfigure[MSE]{\label{fig:generator_test_ssim}\includegraphics[width=0.49\linewidth]{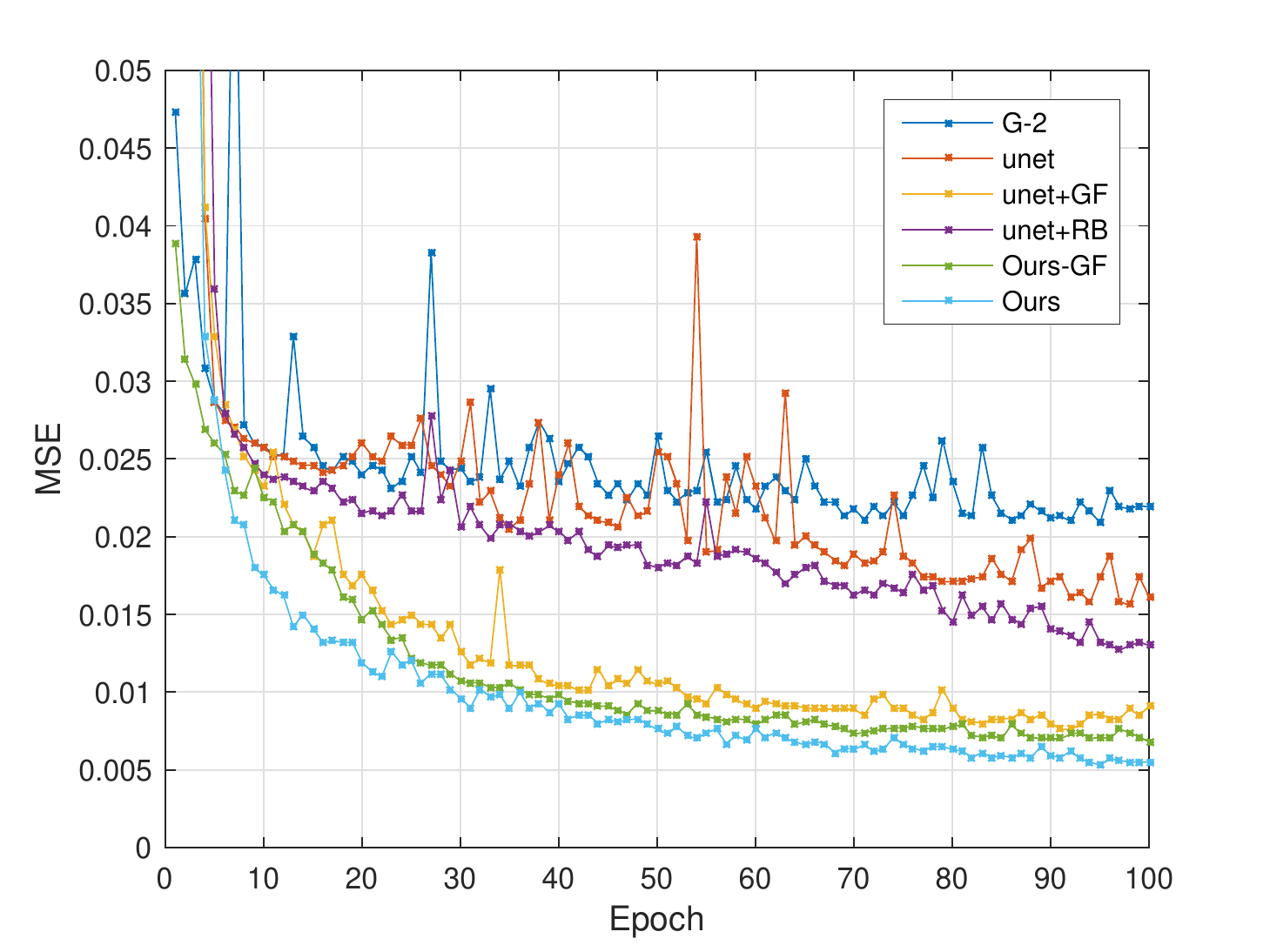}}
	\caption{Comparisons for generator structures. Better zoom in for view. }
	\label{fig:generator_test_psnr_ssim}
\end{figure}

To evaluate the effectiveness of the proposed multi-scale feature extraction and global feature design as well as the proposed generator structure,  we make comparison against the following generator model:  
1) generator used in \cite{li2018emerging}, \cite{lu2019multi} (Two scales of low-level features, here, we term it as G-2 for simplicity).
2)  the basic U-Net as used in \cite{fabbri2018enhancing}.
3) U-Net implemented with residual blocks (RB) in the encoder.
4) U-Net augmented with global features fusion (GF).
5) Our generator without the global feature fusion: denoted as Ours-GF.
6) our proposed generator structure.
Here we construct a small training set  with 512 paired underwater images. The training is conducted with a batch size of 8 for 100 epoch.  The objective for training is to minimize the mean square (MSE) error.  The PNSR and MSE results are reported in table \ref{tab:generator} and the training process is shown in figures \ref{fig:generator_test_psnr_ssim}.
\begin{table}[h]
	\centering
\footnotesize
	\caption{Comparison and evaluation of different generator structures.}
	\label{tab:generator}
	\begin{tabular}{c|cccccc} \hline 
		& G-2& U-Net & UNet+RB & UNet+GF & Ours-GF& Ours   \\ \hline \hline
		PSNR                  & 16.78 & 18.11 & 18.85    & 20.46  &   21.82  & \textbf{22.60 } \\
		MSE                  &  0.022 & 0.016 & 0.013    & 0.0091 & 0.0068  & \textbf{0.0055} \\ \hline
	\end{tabular}
\end{table}

From table \ref{tab:generator}, we can have the following findings. 
1)  The number of the scales of feature matters. The comparison among G-2 (2 scales of features), U-Net (5 scales of features) and Ours-GF (8 scales of features) shows that multi-scale  features are beneficial for  image-to-image translation learning.  
2) Residual blocks in encoder improves the learning ability.
3) Global feature fusion works as well. Overall, our proposed structure can achieve the highest PSNR,  smallest MSE, which demonstrate good learning ability and the effectiveness of the proposed generator structure. 

\section{Conclusions}
\label{section:conclusion}
In this paper, we propose a generic multi-level features extraction and fusion network (MLFcGAN) under the framework of conditional GAN  for underwater image color correction. Extensive experiment results demonstrate that by embedding the high-level information with low-level knowledge at multiple scales,  MLFcGAN posses better learning ability. Our method can effectively restore the underwater images color with fine details and alleviate the unwanted artifacts, which outperform the state of the art both subjectively and objectively.  Furthermore, as feature aggregation is fundamental in solving computer vision  tasks via deep learning,  thus our  strategy could also be explored to other computer vision topics such as segmentation, salient object detection.
{
\bibliographystyle{unsrt}
\bibliography{bmvcbib}
}

\end{document}